\documentclass[conference]{IEEEtran}

\ifCLASSINFOpdf
\else
\fi
\usepackage{amsmath}
\usepackage{comment} %
\usepackage{cite}
\usepackage{graphicx}
\usepackage{epsfig,graphics,subfigure,psfrag,amsmath,amssymb}
\usepackage{amsfonts}
\usepackage{slashbox}
\usepackage{multirow}
\usepackage{amssymb}
\usepackage{algorithm}
\usepackage{algorithmic}
\usepackage{hyperref}
\usepackage{url}
\usepackage{breakurl}
\usepackage{bm}
\usepackage{graphicx}
\usepackage{mathrsfs}
\usepackage{amssymb}
\usepackage{makecell}

\begin{document}
%
\title{Beam Management for Millimeter Wave Beamspace MU-MIMO Systems}

\author{\IEEEauthorblockN{Qing Xue and Xuming Fang}
\IEEEauthorblockA{Key Lab of Information Coding \& Transmission\\
Southwest Jiaotong University, Chengdu 610031, China\\
Email: shdlxxq5460@my.swjtu.edu.cn, xmfang@swjtu.edu.cn}
\and
\IEEEauthorblockN{Ming Xiao}
\IEEEauthorblockA{Communication Theory Department\\
Royal Institute of Technology, Stockholm 100 44, Sweden\\
Email: mingx@kth.se}
}

\maketitle

\begin{abstract}
Millimeter wave (mmWave) communication has attracted increasing attention as a promising technology for 5G networks. One of the key architectural features of mmWave is the use of massive antenna arrays at both the transmitter and the receiver sides. Therefore, by employing directional beamforming (BF), both mmWave base stations (MBSs) and mmWave users (MUEs) are capable of supporting multi-beam simultaneous transmissions. However, most researches have only considered a single beam, which means that they do not make full potential of mmWave. In this context, in order to improve the performance of short-range indoor mmWave networks with multiple reflections, we investigate the challenges and potential solutions of downlink multi-user multi-beam transmission, which can be described as a high-dimensional (i.e., beamspace) multi-user multiple-input multiple-output (MU-MIMO) technique, including multi-user BF training, simultaneous users' grouping, and multi-user multi-beam power allocation. Furthermore, we present the theoretical and numerical results to demonstrate that beamspace MU-MIMO compared with single beam transmission can largely improve the rate performance of mmWave systems.
\end{abstract}


\IEEEpeerreviewmaketitle

\section{Introduction}
According to Cisco forecast, global mobile data traffic will increase sevenfold between 2016 and 2021 \cite{Cisco-forecasts}. Recent researches showed that mmWave communications, operating in 30-300 GHz bands, are promising technologies for meeting the explosive growth of mobile data demand. Compared with existing microwave systems, mmWave systems are faced with two major challenges: severe propagation loss and sensitivity to blockage. To compensate for high propagation loss, directional BF has been widely used as an essential technique to form a highly directional beam pattern with large antenna gain. Thanks to the short wavelengths of mmWave radios ranging from 10 mm to 1 mm, massive antenna arrays can be packed into the limited dimensions of mmWave transceivers. Therefore, with directional BF, it is possible to form multiple beams at both mmWave transmitter and receiver sides in mobile networks. That is, mmWave systems are in fact able to provide high-dimensional MIMO operations \cite{Deconstructing,Beamspace-MIMO-for-Millimeter-Wave,Beamspace-MIMO-Channel-Modeling} and can realize spatial spectrum reuse at close distance \cite{Low-RF-Complexity}. However, most current work does not make full potential of mmWave. For instance, the work in \cite{Multiple-Sector-ID-Capture,Coverage-and-Rate,Beam-searching,Millimeter-Wave-Beam-Alignment} was focused on single beam transmission scenarios, and the work in \cite{Low-RF-Complexity,Nonorthogonal-Beams,Beamspace-MIMO-for-high-dimensional,Near-Optimal-Beamspace} considered the scenarios where only the transmitter side was operating with multiple beams. Moreover, since mmWave radios have limited ability to diffract around obstacles (e.g., human body), the connection between each pair of transmitter and receiver is vulnerable to blockage events.

In this context, aiming at increasing the achievable rate and maintaining connectivity of mmWave mobile networks, we investigated the challenges and potential solutions (including multi-beam selection, cooperative beam tracking, multi-beam power allocation, and synchronization) associated with single-user multi-beam simultaneous transmissions (i.e., beamspace SU-MIMO) in \cite{Beamspace-SU-MIMO-for-Future}. It is worth mentioning that the proposed scheme is only applicable to the short-range scenarios with multiple NLOS paths (e.g., first or second order reflections from floor and/or ceiling in indoor scenarios). In order to further enhance the performance of mmWave systems, we extend our previous work to multi-user scenarios, namely beamspace MU-MIMO, on the basis of existing research results. To the best of our knowledge, there has been no work on this issue. Since the communication environment with multi-user is more complex than that with single user, not only need we to further expand the strategies proposed in \cite{Beamspace-SU-MIMO-for-Future}, but also will we face some new challenges for implementing beamspace MU-MIMO. For instance, due to the transmit beams selected by different MUEs may be (partially) overlapped, the inter-user interference should be seriously considered in beamspace MU-MIMO. This study mainly focuses on the issues of multi-user BF training, simultaneous users' grouping, and multi-user multi-beam power allocation in beamspace MU-MIMO.

The rest of the paper is organized as follows. In Section II, the network model and the basic idea of beamspace MU-MIMO are introduced. Section III first describes muti-user BF training and then proposes a multi-user grouping mechanism. In Section IV, the potential solutions of power allocation for beamspace MU-MIMO are presented and analyzed. Section V shows some numerical results to evaluate the proposed scheme. Finally, Section VI concludes the paper.

\section{System Overview}
In this study, we consider a short-range indoor mmWave network with one reference MBS and $U_{\rm{total}}$ sparsely distributed MUEs. Let $\mathbb{R}$ denote the set of these MUEs. Meanwhile, both the MBS and MUEs are equipped with massive antenna arrays. Thus, with directional BF and space division technique, they are capable of supporting multiple orthogonal beams simultaneously and can realize spectrum reuse, as illustrated in Fig.1. Let $b_{\max }^{\rm{MBS}}$ and $b_{\max }^u$ denote the maximum number of beams that the MBS and MUE $u$ ($u \in \mathbb{R}$) can form, respectively, we generally have $b_{\max }^{\rm{MBS}}  \ge b_{\max }^u$. Let $\mathbb{Q}$ ($\mathbb{Q} \subseteq \mathbb{R}$) be the set of MUEs served simultaneously by the MBS and $U$ be the number of MUEs in $\mathbb{Q}$, we have $1 \le U \le b_{\max }^{\rm{MBS}}$. Furthermore, supposing that $b$ and $b_u$ are the number of operating beams of the MBS and MUE $u$ in actual transmissions, respectively, and considering that the transmit and receive beams are used in pairs in mmWave networks, we have $b = \sum\limits_{u \in \mathbb{Q}} {b_u }  \le b_{\max }^{\rm{MBS}}$, where $1 \le b_u  \le b_{\max }^u$.

The multi-user multi-beam simultaneous transmission scheme investigated in this study can be described as beamspace MU-MIMO defined as Definition 1. Fig. 1 shows an example of beamspace MU-MIMO in two-dimensional (2D) perspective. Note that the analysis is also applicable to three-dimensional (3D) mode. For ease of analysis, similar to \cite{Nonorthogonal-Beams,Beamspace-SU-MIMO-for-Future,A-MAC-Layer-Perspective}, we replace the MBS with $U$ virtual MBSs (vMBSs) located at the same position. Each vMBS serves different MUEs with different transmit beam sets. Moreover, when $1 < b_u  \le b_{\max }^u$, the transmission mode between MUE $u$ and its corresponding vMBS is beamspace SU-MIMO (e.g., for MUE1 and MUE3 in Fig. 1) and it is beamspace SU-SISO when $b_u = 1$ (e.g., for MUE2). In this context, beamspace MU-MIMO can be defined as a set of beamspace SU-MIMO and/or SU-SISO technologies with space division technique.

\emph{Definition 1} (\emph{Beamspace MU-MIMO}): The beamspace MU-MIMO is defined as an mmWave communication mode that an MBS with multiple orthogonal beams can transmit simultaneously to a set of MUEs, where each MUE is with one or more operating beams. That is, denoting $\mathbb{Q}$ as the set of MUEs, $b$ and $b_u$ as the number of transmitting and receiving beams of the MBS and MUE $u$ ($u \in \mathbb{Q}$), respectively, the multi-user multi-beam simultaneous transmissions can be termed as $N \times N_U$ beamspace MU-MIMO, where $U$ is the number of MUEs in $\mathbb{Q}$, $N$ is the total number of transmitting and receiving (T-R) beam pairs between the MBS and the simultaneous transmitting MUEs, $1 \le N \le \min \left\{ {b,\sum\limits_{u \in \mathbb{Q}} {b_u } } \right\}$.

In order to implement beamspace MU-MIMO and, meanwhile, to achieve optimal system performance, we face many challenges as below.

\hangafter 1
\hangindent 1.2em
\noindent
1) Multi-user BF training: Since only one transmit/receive direction's link quality can be detected at a time in traditional BF training (e.g., in 802.11ad/ay), the efficiency of the optimal beam selection is generally very low. It means that the existing beam selection solutions are not entirely applicable to beamspace MU-MIMO. This study utilizes the capability of supporting multiple beams both at the MBS and MUEs to detect the quality of multiple links simultaneously, and thus to increase the efficiency of multi-beam selection for beamspace MU-MIMO.

\hangafter 1
\hangindent 1.2em
\noindent
2) Inter-user interference coordination: The best transmit beam sets selected by different MUEs may be (partially) overlapped, e.g., in Fig. 1, one NLOS link for MUE1 is in conflict with the LOS link for MUE2 over the transmit beam. And the inter-user interference will be severe in this case. To address this issue, we can switch some MUEs' initial selected T-R beam pairs with conflicting beams to a suitable candidate (if available), or assign the MUEs with the same conflicting transmit beam to different groups. MUEs in different groups will be served in time division manner.

\hangafter 1
\hangindent 1.2em
\noindent
3) Power allocation: Considering that the transmission performance of different links for different MUEs may vary widely, the appropriate power allocation strategies should be seriously considered to maximize the achievable rate of beamspace MU-MIMO.

\begin{figure}[t]
  \begin{center}
    \scalebox{0.6}[0.6]{\includegraphics{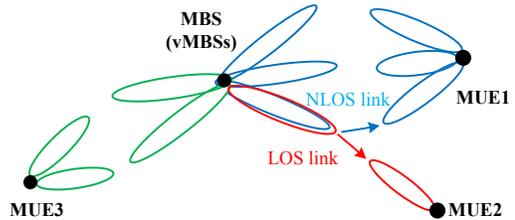}}
    \caption{An example of 2D view of beamspace MU-MIMO ($U = 3$). Note that, in order to make the figure clear, we do not show the side lobes here.}
    \label{fig:1}
  \end{center}
\end{figure}

\section{Beam/User Management for Beamspace MU-MIMO}
In this section, we first give an efficient multi-user BF training mechanism for beamspace MU-MIMO. It is worth mentioning that the BF training (or beam steering) operations are designed to determine the best T-R beam pair set $\mathbb{N}_{\rm{pair}}^u$ ($u \in \mathbb{R}$) that best matches the LOS path and/or NLOS paths between a vMBS and its corresponding MUE, hereafter called vMBS-MUE. In this study, after the successful completion of BF training, directional BF is established. Moreover, to guarantee the network performance, we further adjust the initial selected T-R beam pair sets by analyzing inter-user interference. Table I summarizes the main notations used throughout the paper.

\begin{table}[!h]
\centering
\caption{Summary of Main Notations.}
\begin{tabular}{!{\vrule width0.6pt}c!{\vrule width0.6pt}c!{\vrule width0.6pt}}
\Xhline{0.6pt}
\textbf{Symbol} & \textbf{Definition}\\        
\Xhline{0.6pt}
\Xhline{0.6pt}
$\mathbb{R}$	&  The set of MUEs within the coverage of the MBS\\
\Xhline{0.6pt}
$\mathbb{Q}$	&  The set of MUEs served simultaneously ($\mathbb{Q} \subseteq \mathbb{R}$)\\
\Xhline{0.6pt}
$U$	&  Number of MUEs in $\mathbb{Q}$\\
\Xhline{0.6pt}
$b_{\max }^{\rm{MBS}}$ & Maximum number of simultaneous transmit beams at MBS\\    
\Xhline{0.6pt}
$b_{\max }^u$ & Maximum number of simultaneous receive beams at MUE $u$\\
\Xhline{0.6pt}
$b$	& Number of the operating beams at the MBS\\    
\Xhline{0.6pt}
$b_u$ & Number of the operating beams at MUE $u$ ($u \in \mathbb{Q}$)\\
\Xhline{0.6pt}
$\mathbb{N}_{\rm{TX}}^u$	& Best transmit beam set of MUE $u$ ($u \in \mathbb{R}$)\\
\Xhline{0.6pt}
$\mathbb{N}_{\rm{RX}}^u$  & Best receive beam set of MUE $u$ ($u \in \mathbb{R}$)\\
\Xhline{0.6pt}
$\mathbb{N}_{\rm{pair}}^u$	&  Best T-R beam pair set of vMBS-MUE $u$ ($u \in \mathbb{R}$)\\
\Xhline{0.6pt}
$\mathbb{N}_{\rm{cd}}^u$	&  Candidate T-R beam pair set of vMBS-MUE $u$ ($u \in \mathbb{R}$)\\
\Xhline{0.6pt}
$\mathbb{N}^u$	&  Operating T-R beam pair set of vMBS-MUE $u$ ($u \in \mathbb{Q}$)\\
\Xhline{0.6pt}
$\mathbb{C}_m$	&  The set of MUEs with conflicting beam $m$ ($\mathbb{C}_m \subseteq \mathbb{R}$)\\
\Xhline{0.6pt}
$\eta$	& Threshold of SNR (or SINR)\\
\Xhline{0.6pt}
$\xi _t$	& Transmitting beamwidth\\
\Xhline{0.6pt}
$\xi _r$	& Receiving beamwidth\\
\Xhline{0.6pt}
$R_i^u$	&  Transmission distance of link $i$ for MUE $u$ ($i \in \mathbb{N}^u$)\\
\Xhline{0.6pt}
\end{tabular}
\end{table}

\subsection{Multi-user BF Training}
The multi-user BF training mechanism in this study aims at improving the downlink performance of beamspace MU-MIMO. The corresponding strategies for uplink transmission are left as our future work. For ease of illustration, we divide the region of the MBS/MUEs into a number of transmit/receive sectors (i.e., orthogonal beam directions). Meanwhile, we assume that MUEs can distinguish signals received from different beams. The proposed mechanism mainly consists of three phases and, moreover, the conceptual flow of the first two phases is illustrated in Fig. 2. The details are as follows.

\emph{(i) Transmit Training:} In this phase, all MUEs are in the quasi-omni mode and the MBS scans $n_{\rm{tx}}$ transmit sectors' quality simultaneously with $n_{\rm{tx}}$ directional beams. Here, different sectors are scanned by different beams which are mutually orthogonal. Assuming that the total number of transmit sectors is $S_{\rm{MBS}}$, we have
\begin{equation}
1 \le n_{\rm{tx}} \le \min \left\{ {b_{\max }^{{\rm{MBS}}} ,S_{\rm{MBS}} } \right\}.
\end{equation}
Hence, we only need to test $\left\lceil {\frac{{S_{\rm{MBS}} }}{{n_{\rm{tx}} }}} \right\rceil$ times to determine the best transmit beam set $\mathbb{N}_{\rm{TX}}^u$ for MUE $u$ ($\forall u \in \mathbb{R}$), while the traditional transmit training operating with only one beam at a time is required to test $S_{\rm{MBS}}$ times.

\emph{(ii) Receive Training:} In this phase, it reverses the scanning roles from the transmit training. That is, MUEs detect multiple receive sectors simultaneously with multiple directional beams and the MBS remains in the quasi-omni mode at this time. Similar to transmit training, MUE $u$ ($\forall u \in \mathbb{R}$) can obtain its best receive beam set $\mathbb{N}_{\rm{RX}}^u$ after scaning $\left\lceil {\frac{{S_u }}{{n_{\rm{rx}}^u }}} \right\rceil$ times, where $S_u$ is the total number of MUE $u$'s receive sectors, $n_{\rm{rx}}^u$ is the number of simultaneous scanning beams, and
\begin{equation}
1 \le n_{\rm{rx}}^u \le \min \left\{ {b_{\max }^u ,S_u } \right\}.
\end{equation}
Since the number of simultaneous receive beams supported by each MUE may be different, the number of tests required for completing their respective receive training will also be different. Supposing that $S_u = S_{\rm{MUE}}$ for $\forall u \in \mathbb{R}$, the number of tests required to complete multi-user receive training is $\left\lceil {\frac{{S_{\rm{MUE}} }}{{\mathop {\min }\limits_{u \in \mathbb{R}} n_{\rm{rx}}^u }}} \right\rceil$.

\emph{(iii) Beam Combining:} We first test the transmit and receive beams in $\mathbb{N}_{\rm{TX}}^u$ and $\mathbb{N}_{\rm{RX}}^u$ in pairwise combinations to get multiple T-R beam pair candidates which meet certain communication requirements, e.g., ${\rm{SNR}} \ge \eta$, where $\eta$ is a given threshold. Then, by adopting the multi-beam combination selection algorithm proposed in \cite{Beamspace-SU-MIMO-for-Future}, we can determine $\mathbb{N}_{\rm{pair}}^u$ ($\forall u \in \mathbb{R}$) in which there are $b_u$ T-R beam pair candidates, $b_u  \le b_{\max }^u$. Furthermore, the link quality of each candidate should meet ${\rm{SINR}}_{u,i}  \ge \eta$ ($\forall i \in \mathbb{N}_{\rm{pair}}^u$) when multiple links are transmitted simultaneously. Note that MUEs make the decision independently in this phase and they may obtain several alternative (or sub-optimal) T-R beam pair sets.

\begin{figure}[t]
  \begin{center}
    \scalebox{0.6}[0.6]{\includegraphics{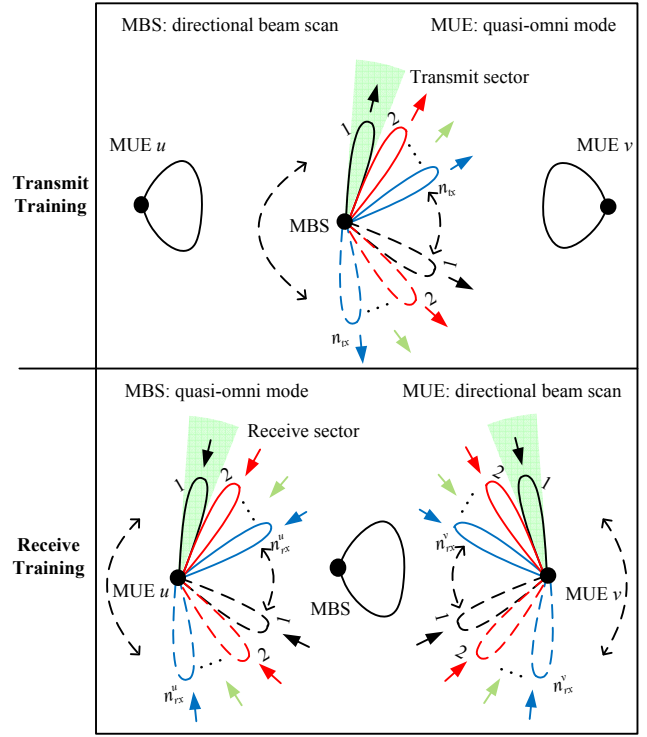}}
    \caption{Illustration of transmit and receive training for downlink beamspace MU-MIMO, given that $U_{\rm{total}} = 2$. The beams drawn with solid lines are operated concurrently and it is the same to that drawn with dotted lines.}
    \label{fig:1}
  \end{center}
\end{figure}

\subsection{Simultaneous MUEs' Grouping}
Considering the diversity and finiteness of the number of simultaneous operating beams that can be supported by the MBS and MUEs, the MBS is generally unable to serve all MUEs in its coverage simultaneously. Moreover, the system performance of beamspace MU-MIMO is various for different combinations of simultaneous MUEs. To ensure the performance, this subsection is devoted to grouping simultaneous MUEs through the analysis of inter-user interference.

Since the decision of multi-beam selection for each MUE is relatively independent, the transmit beams selected by them may be (partially) overlapped. We assume that one beam can serve only one MUE at the same time. To avoid beam conflicts, we need to adjust or re-select $\mathbb{N}_{\rm{pair}}^u$ ($\forall u \in \mathbb{R}$), e.g., by beam switching. After that, we can proceed with the selection of simultaneous MUEs as described in Algorithm 1, the main idea of which can be outlined as follows:

\hangafter 1
\hangindent 1.2em
\noindent
$\bullet$ Denoting $\mathbb{C}_m$ as the set of MUEs with conflicting transmit beam $m$, the MBS gives priority to MUE $s$ ($s \in \mathbb{C}_m$) which satisfies ${\rm{SINR}}_{s,m}  = \mathop {\max }\limits_{u \in \mathbb{C}_m } {\rm{SINR}}_{u,m} $.

\hangafter 1
\hangindent 1.2em
\noindent
$\bullet$ The MUEs (e.g., MUE $u$, $u \in \mathbb{C}_m \backslash s$), which satisfy ${\rm{SINR}}_{u,m} \ne \mathop {\max }\limits_{i \in \mathbb{N}_{{\rm{pair}}}^u } {\rm{SINR}}_{u,i}$, can switch their initial selected T-R beam pair sets to the best suitable candidates $\mathbb{N}_{{\rm{cd}}}^u$ (if available). Here, \emph{suitable candidate} refers to the alternative T-R beam pair set which satisfies: (1) $\mathbb{N}_{{\rm{cd}}}^u  \cap \mathbb{N}_{{\rm{pair}}}^v = \emptyset$, $\forall v \in \mathbb{Q}\backslash u$; (2) ${\rm{SINR}}_{u,i}  \ge \eta $, $\forall i \in \mathbb{N}_{{\rm{cd}}}^u$. If these MUEs have no suitable candidates, they will be served in time division manner.

\hangafter 1
\hangindent 1.2em
\noindent
$\bullet$ The other MUEs (e.g., MUE $k$, $k \in \mathbb{C}_m \backslash s,u$) should be assigned to different simultaneous MUE groups and will be served in time division manner.

Denoting $\mathbb{N}^u$ as the operating T-R beam pair set of MUE $u$, we have $\mathbb{N}^u = \mathbb{N}_{{\rm{pair}}}^u$ or $\mathbb{N}^u = \mathbb{N}_{{\rm{cd}}}^u$. In this study, the MBS carries out the decision of multi-user grouping and informs MUEs of the decision result. Before each transmission cycle, Algorithm 1 can realize the selection of simultaneous MUEs for beamspace MU-MIMO. Meanwhile, the non-selected MUEs have relative high priorities in the next cycle to ensure fairness.

\begin{algorithm}[t]
\caption{Multi-user grouping}
\label{alg:1}
\begin{algorithmic}[1]
\REQUIRE ~~\\
$\bullet$ the maximum number of transmit beams $b_{\max }^{\rm{MBS}}$;\\
$\bullet$ the best T-R beam pair sets $\mathbb{N}_{\rm{pair}}^u$ ($\forall u \in \mathbb{R}$);
\STATE Initialize $\mathbb{Q} = \emptyset$;
\STATE Compare $\mathbb{N}_{\rm{pair}}^u $ and $\mathbb{N}_{\rm{pair}}^v$ ($\forall u,v \in \mathbb{R}$, $u \ne v$);
\STATE Record the MUEs without beam conflicts into $\mathbb{Q}_1$;
\IF{$\mathbb{Q}_1 = \mathbb{R}$ or $b = \sum\limits_{u \in \mathbb{Q}_1} {b_u }  > b_{\max }^{{\rm{MBS}}}$}
\STATE Rank MUEs in $\mathbb{Q}_1$ in decreasing order according to the average link quality $\overline {{\rm{SINR}}}$, e.g., $\overline {{\rm{SINR}}}_u  = \frac{{\sum\nolimits_{i \in \mathbb{N}_{\rm{pair}}^u } {{\rm{SINR}}_{u,i} } }}{{b_u }}$;
\STATE Record the first $U$ MUEs into $\mathbb{Q}$ which satisfies $b = \sum\limits_{u \in \mathbb{Q}} {b_u }  \le b_{\max }^{{\rm{MBS}}}$;
\ELSE
\STATE Select an MUE (e.g., MUE $s$) in each set of MUEs with conflicting beam (e.g., $\mathbb{C}_m$), which satisfies ${\rm{SINR}}_{s,m}  = \mathop {\max }\limits_{u \in \mathbb{C}_m } {\rm{SINR}}_{u,m} $, and record them into $\mathbb{Q}_1$;
\STATE Record the MUEs who can switch to a suitable candidate T-R beam pair set in ($\mathbb{R} - \mathbb{Q}_1$) into $\mathbb{Q}_1$;
\STATE Rank MUEs in $\mathbb{Q}_1$ and Record the first $U$ MUEs into $\mathbb{Q}$ as in step 5 and step 6, respectively;
\ENDIF
\ENSURE the set of simultaneous MUEs $\mathbb{Q}$
\end{algorithmic}
\end{algorithm}

\section{Power Allocation for Beamspace MU-MIMO}
Since the quality of different links may vary widely among simultaneous MUEs, we should make reasonable power allocation for beamspace MU-MIMO in order to maximize its achievable rate. The NLOS links in this study are assumed to be first order reflections, because mmWave signals are generally negligible after high-order reflections and the actual transmission paths of them are unpredictable. For tractability of the analysis, we approximate the actual antenna pattern by an ideal sectored antenna model \cite{Coverage-and-Rate,Transmission-capacity}. The directivity gain can be expressed as \cite{Beam-searching,A-MAC-Layer-Perspective}
\begin{equation}
G \left( {\xi} \right) = \left\{ {\begin{array}{*{20}c}
   {\frac{{2\pi  - \left( {2\pi  - \xi } \right)z}}{{\xi}},} & \rm{in\,the\,main\,lobe,}  \\
   {z,} & \rm{in\,side\,lobes,}  \\
\end{array}} \right.
\end{equation}
where $\xi$ is the operating beamwidth and $z$ is the average gain of side lobes, $0 \le z < 1$. Furthermore, the path loss of mmWave can be modeled as \cite{Channel-Models}
\begin{equation}
L\left( {R} \right)\left[ {\rm{dB}} \right] = A + 20\log _{10} \left( {f_c } \right) + 10n\log _{10} \left( R \right),
\end{equation}
where $f_c$ is the carrier frequency in GHz, $R$ is transmission distance in km, $A$ is the attenuation value, and $n$ is the path loss exponent. Since the T-R beam pairs for beamspace MU-MIMO are mutually orthogonal (i.e., the main lobes are non-overlapping), we assume that the inter-beam interference is mainly caused by side lobes. Therefore, the SINR of link $i$ for MUE $u$ ($u \in \mathbb{Q}$) is
\begin{equation}
{\rm{SINR}}_{u,i} \left[ {{\rm{dB}}} \right] = 10\log _{10} \frac{{P_t^i  \cdot g \left( {\xi _t^{u,i} } \right) \cdot g \left( {\xi _r^{u,i} } \right) \cdot \frac{1}{{L\left( {R_i^u } \right)}}}}{{P_N  + \sum\limits_{j \in \mathbb{M}} {{P_t^j  \cdot z \cdot g \left( {\xi _r^{u,i} } \right) \cdot \frac{1}{{L\left( {R_i^u } \right)}}} } }},
\end{equation}
where $P_t^i$ is the transmitted power; $g \left( {\xi _t^{u,i} } \right) = \frac{{2\pi  - \left( {2\pi  - \xi _t^{u,i} } \right)z}}{{\xi _t^{u,i} }}$ and $g \left( {\xi _r^{u,i} } \right) = \frac{{2\pi  - \left( {2\pi  - \xi _r^{u,i} } \right)z}}{{\xi _r^{u,i} }}$ are the average main lobe gains of transmit and receive beams, respectively; $P_N$ is the thermal noise power; $\mathbb{M} = \left( {\mathbb{N}^u \backslash i} \right) \cup \left( {\mathop  \cup \limits_{v \in \mathbb{Q}\backslash u} \mathbb{N}^v } \right)$. Moreover, the achievable rate of link $i$ can be estimated as ${\rm{Rate}}_{u,i}   = B \cdot \log _2 \left( {1 + {\rm{SINR}}_{u,i} } \right)$ according to Shannon capacity formula, where $B$ is the operating bandwidth.

To maximize the achievable rate of beamspace MU-MIMO, we first collect the variables $\xi _t^{u,i}$, $\xi _r^{u,i}$, and $P_t^i$ ($\forall u \in \mathbb{Q}$, $\forall i \in \mathbb{N}^u$) in vectors ${\bm{\xi _t }}$, ${\bm{\xi _r }}$ and ${\bm{p}}$, respectively, and then formulate the problem under consideration as an optimization problem (P1) given by
\begin{subequations}
\begin{align}
\mathop {\rm{maximize} }\limits_{{\bm{\xi _t }},{\bm{\xi _r}} ,{\bm{p}},\mathbb{Q}}  \quad & {\rm{Rate}} = {\sum\limits_{u \in \mathbb{Q}} {\sum\limits_{i \in \mathbb{N}^u }  {B \cdot \log _2 \left( {1 + {\rm{SINR}}_{u,i} } \right)} } }\\
{\rm{subject\,to}}\quad & \xi _{t,\min }  \le \xi _t^{u,i}  < {2\pi },\\
& \xi _{r,\min }  \le \xi _r^{u,i}  < {2\pi },\\
& {1 \le U \le b_{\max }^{\rm{MBS}} },\\
& 0 \le P_t^i  \le p_{\max } ,\\
& 0 < {\sum\limits_{u \in \mathbb{Q}} {\sum\limits_{i \in \mathbb{N}^u }  {P_t^i} } }  \le P_{\max },
\end{align}
\end{subequations}
where $\xi _{t,\rm{min} }$ and $\xi _{r,\rm{min} }$ are the minimum beamwidth of transmit and receive beams, respectively; $p_{\rm{max}}$ and $P_{\rm{max}}$ are the maximum transmission power of each transmit beam and the MBS, respectively. Note that function arguments have been discarded for notational simplicity. Considering the simplest scenario with pencil beams, i.e., $z \ll 1$, we can neglect the inter-beam interference and optimize the operating beamwidth for each link individually. That is, the optimal beamwidth of transmit and receive beams are $\left( {\xi _t^{u,i} } \right)^ * = \xi _{t,\min }$ and $\left( {\xi _r^{u,i} } \right)^ * = \xi _{r,\min }$, respectively. Hereafter, the optimized parameters are identified by the "$*$" on the upper right corner. Meanwhile, the SINR expression formulated in Eq. (5) reduces to SNR according to
\begin{equation}
{\rm{SNR}}_{u,i}^* = \frac{{\left( {P_t^i } \right)^* \cdot \frac{{2\pi }}{{\xi _{t,\min } }} \cdot \frac{{2\pi }}{{\xi _{r,\min } }} \cdot \frac{1}{{L\left( {R_i^u } \right)}}}}{{P_N }}.
\end{equation}

As P1 is difficult to obtain its optimal solution, we investigate two low complexity and easy to implement solutions for multi-user multi-beam power allocation to suboptimally address P1 that with pencil beams.

$\star$ \emph{Average Power Allocation (APA):} Each link's transmission power is the same without considering the difference of link quality, i.e., $\left( {P_t^i } \right)_{\rm{APA}}^*  = \frac{{P_{\max } }}{{\sum\limits_{u \in \mathbb{Q}^* } {b_u ^* } }}$, where $\mathbb{Q}^*$ and $b_u ^*$ can be obtained by Algorithm 2.

$\star$ \emph{Priority Power Allocation (PPA):} Considering the quality of different links may vary widely, we give priority to optimize the transmission power of the links that with high link quality to address P1. Furthermore, this solution includes the following two cases.

1) Considering fairness: When the fairness of power allocation among simultaneous MUEs is taken into account, we will give priority to optimize the best link for each MUE (e.g., link $\ell$ for MUE $u$), i.e., $\left( {P_t^{u,\ell } } \right)_{\rm{FP}}^*  = p_{\max }$. Then, we employ APA to allocate power for other links (e.g., link $i$), i.e., $\left( {P_t^{u,i} } \right)_{\rm{FP}}^*  = \frac{{P_{\max }  - U^*  \cdot p_{\max } }}{{\sum\limits_{u \in \mathbb{Q}^* } {\left( {b_u ^*  - 1} \right)} }}$, $\forall i \in \left( {\mathbb{N}^u } \right)^* \backslash \ell$, where $U^*$ is the number of MUEs in $\mathbb{Q}^*$.

2) Without considering fairness: We rank all the links in $\mathbb{N}$ in decreasing order according to the link quality, where $\mathbb{N} = \mathop  \cup \limits_{u \in \mathbb{Q}} \mathbb{N}^u$. Denoting $\mathbb{N}_{\rm{OFP}}^*$ as the set of the first $\left\lfloor {\frac{{P_{\max } }}{{p_{\max } }}} \right\rfloor$ links in $\mathbb{N}$, we have $\left( {P_t^i } \right)_{\rm{OFP}}^*  = p_{\max }$ for $\forall i \in \mathbb{N}_{\rm{OFP}}^*$. Further, if the rest of the power $p = P_{\max }  - \left\lfloor {\frac{{P_{\max } }}{{p_{\max } }}} \right\rfloor  \cdot p_{\max }$ can meet the communication requirements of a link in $\left( \mathbb{N} - \mathbb{N}_{\rm{OFP}}^* \right)$, we have $U^*  = \left\lceil {\frac{{P_{\max } }}{{p_{\max } }}} \right\rceil$; Otherwise, $U^*  = \left\lfloor {\frac{{P_{\max } }}{{p_{\max } }}} \right\rfloor$.

Substituting the optimized parameters into Eq. (6) and (7), we can easily obtain the maximum achievable rate with APA and with PPA, respectively. Since the length of the paper is limited, we do not give these results here.

\begin{algorithm}[t]
\caption{$b_u$ ($\forall u \in \mathbb{Q}$) Optimization for APA}
\label{alg:1}
\begin{algorithmic}[1]
\REQUIRE ~~\\
$\bullet$ the set of simultaneous MUEs $\mathbb{Q}$;\\
$\bullet$ the operating T-R beam pair sets $\mathbb{N}^u$ for $\forall u \in \mathbb{Q}$;
\STATE $\overline P _t  = \frac{{P_{\max } }}{{\sum\limits_{u \in \mathbb{Q}} {b_u } }}$;
\IF{$\overline P _t > p_{\max }$}
\STATE Let $\overline P _t  = p_{\max }$;
\ENDIF
\STATE ${\rm{SNR}}_{u,i} = \frac{{\overline P _t \cdot \frac{{2\pi }}{{\xi _{t,\min } }} \cdot \frac{{2\pi }}{{\xi _{r,\min } }} \cdot \frac{1}{{L\left( {R_i^u } \right)}}}}{{P_N }}$ for $\forall u \in \mathbb{Q}$, $\forall i \in \mathbb{N}^u$;
\IF{$\mathop {\min }\limits_{u \in \mathbb{Q}, i \in \mathbb{N}^u } {\rm{SNR}}_{u,i}  < \eta$}
\STATE Remove link $j$ from $\mathbb{N}^s$ ($s \in \mathbb{Q}$), the link satisfies ${\rm{SNR}}_{s,j} = \mathop {\min }\limits_{u \in \mathbb{Q}, i \in \mathbb{N}^u } {\rm{SNR}}_{u,i}$;
\STATE $b_s = b_s - 1$;
\IF{$b_s = 0$}
\STATE Remove MUE $s$ from $\mathbb{Q}$;
\ENDIF
\STATE Go to step 1;
\ELSE
\STATE $\mathbb{Q}^* = \mathbb{Q}$;
\STATE $\left( \mathbb{N}^u \right)^* = \mathbb{N}^u$ and $b_u^* = b_u$ for $\forall u \in \mathbb{Q}^*$;
\ENDIF
\ENSURE $b_u^*$ ($\forall u \in \mathbb{Q}^*$)
\end{algorithmic}
\end{algorithm}

\begin{table}[!h]
\centering
\caption{Simulation Parameters.}
\begin{tabular}{!{\vrule width0.6pt}c!{\vrule width0.6pt}c!{\vrule width0.6pt}}
\Xhline{0.6pt}
\textbf{Parameters} & \textbf{Values}\\
\Xhline{0.6pt}
\Xhline{0.6pt}
Carrier frequency, $f_c$	& 60GHz\\
\Xhline{0.6pt}
Bandwidth, $B$	& 1.5GHz\\
\Xhline{0.6pt}
Maximum transmit power of MBS, $P_{\max }$	& 10dBm\\
\Xhline{0.6pt}
Maximum power of transmit beams, $p_{\max }$	& 3dBm\\
\Xhline{0.6pt}
Maximum number of transmit beams, $b_{\max }^{\rm{MBS}}$	& 10\\
\Xhline{0.6pt}
Maximum number of receive beams, $b_{\max }^{\rm{MUE}}$	& 3\\
\Xhline{0.6pt}
\multirow{2}{*}{Attenuation value, $A$}
& $A_{{\rm{LOS}}} = 32.5$;\\
& $A_{{\rm{NLOS}}} = 45.5$\\
\Xhline{0.6pt}
\multirow{2}{*}{Path loss exponent, $n$}	
& $n_{{\rm{LOS}}} = 2.0$;\\
& $n_{{\rm{NLOS}}} = 1.4$\\
\Xhline{0.6pt}
\multirow{2}{*}{Transmission distance, $R$}	
& $R_{{\rm{LOS}}} = 7m$;\\
& $R_{{\rm{NLOS}}} = 10m$\\
\Xhline{0.6pt}
Noise figure, $NF$	& 6dB\\
\Xhline{0.6pt}
\end{tabular}
\end{table}

\section{Performance Evaluation}
This section presents some numerical results on the performance of beamspace MU-MIMO. The objective of this work is two-fold: (i) to verify the effectiveness of the proposed multi-user BF training mechanism; (ii) to compare and analyze the performance of APA and PPA. To simplify simulations, we assume that $\xi_t^{u,i} = \xi_t$, $\xi_r^{u,i} = \xi_r$, and $b_{\max}^u = b_{\max }^{\rm{MUE}}$, for $\forall i \in \mathbb{N}^u$, $\forall u \in \mathbb{Q}$. Moreover, we consider a short-range indoor mmWave network with $R_i^u = R_{\rm{LOS}}$ for LOS links and $R_i^u = R_{\rm{NLOS}}$ for NLOS links. Table II summarizes the detailed simulation parameters. In addition, at a standard temperature of $17\,^{\circ}\mathrm{C}$, we let $P_N\left[ \rm{{dB}} \right]=  - 174\left[\rm{ {{{dBm} \mathord{\left/
 {\vphantom {{dBm} {Hz}}} \right.
 \kern-\nulldelimiterspace} {Hz}}}} \right] + 10\log _{10} \left( B \right) + NF$, where $NF$ is noise figure in dB.

Fig. 3 shows that the proposed multi-beam transmit training can effectively improve the efficiency of beam selection. For example, when $\xi_t = 10^\circ$, we have $S_{\rm{MBS}} = 36$ and $1 \le n_{\rm{tx}} \le 10$ which can be known from Eq. (1). Hence, if $n_{\rm{tx}} = 5$, to obtain the best transmit beam sets (i.e., $\mathbb{N}_{\rm{TX}}^u$, $\forall u \in \mathbb{Q}$), the MBS only needs to scan $\left\lceil {\frac{{S_{\rm{MBS}} }}{{n_{\rm{tx}} }}} \right\rceil = 8$ times by adopting the proposed solution. However, using the traditional transmit training operating with single beam, the number of scans required is $S_{\rm{MBS}} = 36$. Similarly, we can verify the effectiveness of the proposed multi-beam receive training for selecting the best receive beam sets (i.e., $\mathbb{N}_{\rm{RX}}^u$, $\forall u \in \mathbb{Q}$). Furthermore, the larger the values of $n_{\rm{tx}}$ and $n_{\rm{rx}}$, the more superior the multi-user BF training.

In Fig. 4, we investigate the rate performance of beamspace MU-MIMO with APA and with PPA, respectively. Here we consider the network is with three MUEs, i.e., $U=3$ and, meanwhile, each of them is operating with a LOS link and two NLOS links, i.e., $b = \sum\limits_{u \in \mathbb{Q}} {b_u } = 9$. Moreover, we assume that the quality of each LOS link is better than that of NLOS links. In this context, the rate performance of beamspace MU-MIMO with PPA is the same regardless of whether the fairness of power allocation among simultaneous MUEs is taken into account. Clearly, the performance of PPA is generally better than that of APA for beamspace MU-MIMO. Further, the results indicate that beamspace MU-MIMO compared with beamspace MU-SISO can largely improve the rate performance of mmWave networks. Note that $U>1$ and $b_u = 1$ ($\forall u \in \mathbb{Q}$) in beamspace MU-SISO systems. For example, when $\eta = 10{\rm{dB}}$ shown in Fig. 4(a), we have ${\rm{Rate}}_{\rm{MU-MIMO}}^{\rm{APA}}=120 {\rm{Gbps}}$ and ${\rm{Rate}}_{\rm{MU-MIMO}}^{\rm{PPA}}=210 {\rm{Gbps}}$ while ${\rm{Rate}}_{\rm{MU-SISO}}=49 {\rm{Gbps}}$. To make the results more general, the inter-beam interference caused by side lobes is not ignored in our simulations, i.e., $z \ne 0$. By comparing Fig. 4(a) with Fig. 4(b), we can see that the greater the value of $z$, the greater the interference, and the more obvious the impact on the system performance.

\begin{figure}[t]
  \begin{center}
    \scalebox{0.55}[0.55]{\includegraphics{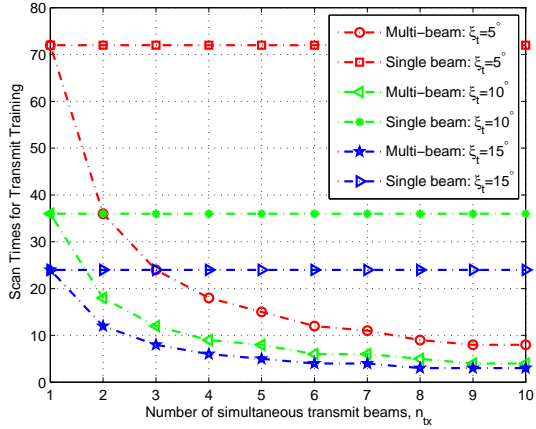}}
    \caption{Performance comparison between the proposed multi-beam transmit training and the traditional solution.}
    \label{fig:1}
  \end{center}
\end{figure}

\begin{figure}[t]
  \begin{center}
    \scalebox{0.55}[0.55]{\includegraphics{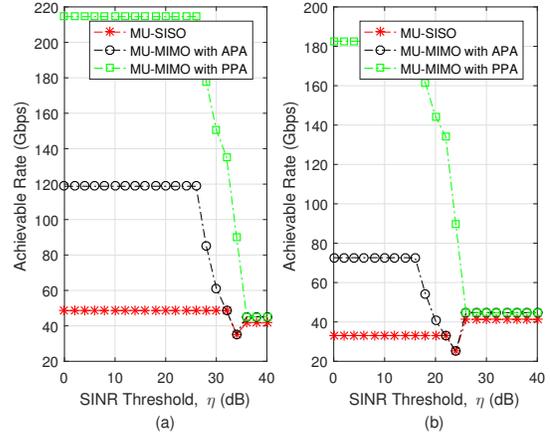}}
    \caption{Achievable rate performance versus SINR threshold $\eta$ for beamspace MU-SISO and MU-MIMO with PPA and with APA, respectively, given that $\xi_t = 10^\circ$, $\xi_r = 15^\circ$, and (a) $z = 0.01$, (b) $z = 0.1$.}
    \label{fig:1}
  \end{center}
\end{figure}

\section{Conclusions}
In this study, in order to further enhance the performance of mmWave networks with multiple reflections, we extended our previous work to multi-user scenario, namely beamspace MU-MIMO, and investigated its challenges and potential solutions for downlink transmission. First, we improved the efficiency of multi-beam selection for beamspace MU-MIMO by utilizing the capability of supporting multiple beams both at the MBS and MUEs. Second, to avoid beam conflicts, we grouped simultaneous served MUEs. Third, we analyzed two low complexity multi-user multi-beam power allocation solutions, i.e., APA and PPA. The numerical results demonstrated that they are very effective to improve the achievable rate of beamspace MU-MIMO. Furthermore, the corresponding strategies for uplink beamspace MU-MIMO will be also our future work.



\bibliographystyle{IEEEtran}
\bibliography{reference}

\end{document}